\documentclass{article}
\usepackage{arxiv}

\AtBeginDocument{%
  \providecommand\BibTeX{{%
    \normalfont B\kern-0.5em{\scshape i\kern-0.25em b}\kern-0.8em\TeX}}}

\usepackage{tikz}
\usepackage{float}
\usepackage{amsmath}
\usetikzlibrary{positioning}
\usetikzlibrary{arrows,arrows.meta}
\usetikzlibrary{decorations.pathreplacing}
\usetikzlibrary{calc}
\usetikzlibrary{arrows,positioning,automata,shadows,fit,shapes}
\usetikzlibrary{trees}

\usepackage[nolist,nohyperlinks]{acronym}
\usepackage{hyperref}
\usepackage{breakurl}

\usepackage{xcolor}
\usepackage{listings,chngcntr}
\usepackage{csquotes}
\usepackage[english]{babel}
\defineshorthand{"~}{\babelhyphen{nobreak}}
\useshorthands{"}

\usepackage{pgfplots}

\usepackage{cleveref}
\makeatletter
\newcommand*{\org@overidelabel}{}
\let\org@overridelabel\@verridelabel
\@ifpackagelater{acronym}{2015/03/21}{
  \renewcommand*{\@verridelabel}[1]{%
    \@bsphack
    \protected@write\@auxout{}{\string\AC@undonewlabel{#1@cref}}%
    \org@overridelabel{#1}%
    \@esphack
  }%
}{
  \renewcommand*{\@verridelabel}[1]{%
    \@bsphack
    \protected@write\@auxout{}{\string\undonewlabel{#1@cref}}%
    \org@overridelabel{#1}%
    \@esphack
  }%
}
\makeatother

\usepackage{amssymb}
\usepackage{pifont}
\newcommand{\cmark}{\ding{51}}%
\newcommand{\xmark}{\ding{55}}%
\newcommand{\omark}{\ding{109}}%
\newcommand*\circled[1]{\tikz[baseline=(char.base)]{%
            \node[shape=circle,fill=black,draw,inner sep=0.1pt] (char) {\footnotesize #1};}}

\newif\ifComments
\Commentstrue

\usepackage{filecontents}

\title{Efficient Intrusion Detection on Low-Performance Industrial IoT Edge Node Devices}

\author{
 Matthias Niedermaier \\
 matthias.niedermaier@hs-augsburg.de \\
 Hochschule Augsburg
 \And
 Martin Striegel \\
 martin.striegel@aisec.fraunhofer.de \\
 Fraunhofer AISEC
 \And
 Felix Sauer \\
 felix.sauer@hs-augsburg.de \\
 Hochschule Augsburg
 \And
 Dominik Merli \\
 dominik.merli@hs-augsburg.de \\
 Hochschule Augsburg
 \And
 Georg Sigl \\
 sigl@tum.de \\
 TU M\"unchen
}

\begin{document}
\maketitle

\begin{abstract}
Communication between sensors, actors and \acp{PLC} in industrial systems moves from two"~wire field buses to \acs{IP}"~based protocols such as Modbus/TCP\@.
This increases the attack surface because the \acs{IP}"~based network is often reachable from everywhere within the company.
Thus, centralized defenses, e.g. at the perimeter of the network do not offer sufficient protection.
Rather, decentralized defenses, where each part of the network protects itself, are needed.
Network \acp{IDS} monitor the network and report suspicious activity.
They usually run on a single host and are not able to capture all events in the network 
and they are associated with a great integration effort.
To bridge this gap, we introduce a method for intrusion detection that combines distributed agents on \ac{IIoT} edge devices with a centralized logging.
In contrast to existing \acp{IDS}, the distributed approach is suitable for industrial low performance microcontrollers.
We demonstrate a \ac{PoC} implementation on a \acs{MCU} running FreeRTOS with LwIP and show the feasibility of our approach in an \ac{IIoT} application.
\end{abstract}


\maketitle

\acresetall
\section{Introduction}
Concepts such as predictive maintenance drive the hunger for more and more data.
Nowadays, edge node devices such as \acp{PLC} and sensors, which provide this data, are usually connected by \acs{IP}"~based protocols~\cite{gupta2010networked}.
This has the advantage that data is accessible from all over the corporate network.
On the other hand, this drastically increases the attack surface as e.g. attackers with access to the network can launch an attack from anywhere inside that network, 
circumventing centralized or perimeter defenses.
Additionally, Internet"~wide search engines such as censys\cite{durumeric2015search} and Shodan\cite{matherly2009shodan} reveal, 
that large numbers of \ac{IIoT} devices are exposed to the Internet without any protection~\cite{mirian2016internet}. 
This allows remote attackers to directly target these devices for example with \ac{DoS} attacks~\cite{niedermaier2018you} or packet injection to send control commands~\cite{beresford2011exploiting} to the victim device.

Due to that, network access control mechanisms are insufficient.
Rather, a defense in depth security concept is needed, in which each networked device needs to be secured individually against network attacks.
\acp{IDS}, which monitor network traffic, can be used to achieve this.

\begin{figure}[htb]
  \centering
  \begin{footnotesize}
\begin{tikzpicture}[node distance=1cm, 
    shorten >= 3pt,shorten <= 3pt,
    auto]    
  \node [circle, draw, fill=white, text width=1.0cm, minimum size=1.5cm,
    inner ysep=0.1cm, text centered] at (1,4) (c1) {Collect Data};    
  \node [circle, draw, fill=white, text width=1.0cm, minimum size=1.5cm,
    inner ysep=0.1cm, text centered] at (3,4) (c2) {Transmit Data};    
  \node [circle, draw, fill=white, text width=1.0cm, minimum size=1.5cm,
    inner ysep=0.1cm, text centered] at (5,4) (c3) {Save Data};    
  \node [circle, draw, fill=white, text width=1.0cm, minimum size=1.5cm,
    inner ysep=0.1cm, text centered] at (7,4) (c4) {Process Data};
    
  \draw [<->, bend angle=45, bend right] (c1.south) 
      to node[below, text width=1.5cm, text centered]{high bandwidth} (c2.south);
  \draw [<->, bend angle=45, bend right] (c2.south) 
      to node[below, text width=1.0cm, text centered]{high storage} (c3.south);
  \draw [<->, bend angle=45, bend right] (c3.south) 
      to node[below, text width=1.0cm, text centered]{high \\ performance} (c4.south);
  
  \node [anchor=center] at (2,5) (en) {Edge node};
  \node [anchor=center] at (6,5) (se) {Server};
     \end{tikzpicture}
  \end{footnotesize}
\caption{Centralized data collection approach with system requirements shown at the bottom.}
\label{fig_classical}
\end{figure}
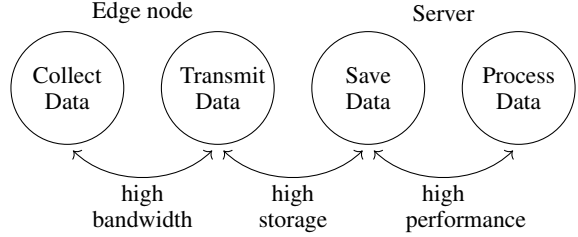

In classical \textit{centralized} monitoring systems, low-performance edge devices forward all information to a high performance system such as a server\footnote{\href{https://www.analog.com/en/technical-articles/intelligence-at-the-edge-part-1-the-edge-node.html}{https://www.analog.com/en/technical-articles/intelligence-at-the-edge-part-1-the-edge-node.html}}.
There, data is stored and processed,
as shown in \Cref{fig_classical}.
However, this approach has several drawbacks.
High bandwidth is required to forward data from sensor nodes to the server.
This places high burden on the network.
Secondly, with more and more sensor nodes, tremendous amounts of data need to be processed at the server, bringing serious scalability issues. 

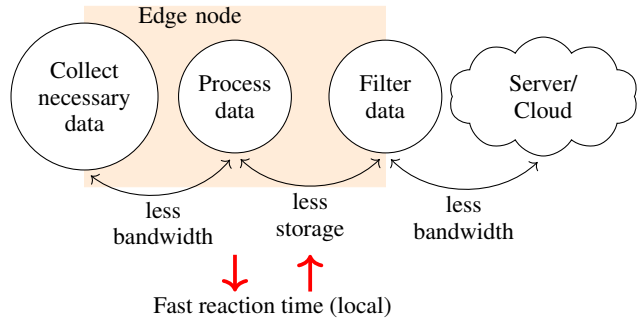
\begin{figure}[htb]
  \centering
  \begin{footnotesize}
\begin{tikzpicture}[node distance=1cm, 
    shorten >= 3pt,shorten <= 3pt,
    auto]    
  
  \filldraw[draw=white, rectangle, color=orange!15!white] 
    (1,-1.2) -- (5,-1.2) -- (5,1.2) -- (1,1.2) -- cycle
    node[color=black, anchor=south west, text width=4cm] (n0) 
    {\hspace*{0.5cm}  Edge node \vspace*{2.0cm}}; 
    
  \node [circle, draw, fill=white, text width=1.4cm, minimum size=1.5cm,
    inner ysep=0.1cm, text centered] at (1,0) (n1) {Collect necessary data};
    
  \node [circle, draw, fill=white, text width=1.0cm, minimum size=1.5cm,
    inner ysep=0.1cm, text centered] at (3,0) (n2) {Process data};
    
  \node [circle, draw, fill=white, text width=1.0cm, minimum size=1.5cm,
    inner ysep=0.1cm, text centered] at (5,0) (n3) {Filter data};
    
  \node [cloud, draw, fill=white,cloud puffs=10,cloud puff arc=120, aspect=2, 
    text width=1.5cm,
    inner ysep=0.0cm, text centered] at (7.1,0) (n4) {Server/\\Cloud};

  \draw [<->, bend angle=45, bend right] (n1.south) 
      to node[below, text width=1.5cm, text centered]{less bandwidth} (n2.south);
  \draw [<->, bend angle=45, bend right] (n2.south) 
      to node[below, text width=1.0cm, text centered]{less storage} (n3.south);
  \draw [<->, bend angle=45, bend right] (n3.south) 
      to node[below, text width=1.5cm, text centered]{less bandwidth} (n4.south);
      
  \draw [->, line width=0.5mm, color=red] (3,-2.0) 
      to node[below]{} (3,-2.7);   
  \draw [<-, line width=0.5mm, color=red] (4,-2.0) 
      to node[below]{} (4,-2.7);   
  \node [anchor=center] at (3.5,-2.8) (cl) {Fast reaction time (local)};
 \end{tikzpicture}
  \end{footnotesize}
\caption{Distributed data collection approach. Preliminary data processing is conducted at the edge devices.}
\label{fig_edge}
\end{figure}

For above reasons, it is beneficial to analyze network data already at the edge device.
As Garcia et al.~\cite{garcia2015edge} note, 
the edge node devices are getting smarter again, because on the one hand, they have enough performance.
On the other hand, this offers advantages in terms of privacy and security.
\Cref{fig_edge} shows a \textit{distributed \ac{IDS}} approach, where every edge node can observe the network, preprocess data and decide autonomously.
While a centralized logging system might still be present, only few data is sent to it.
This has several benefits over the centralized approach:
Bandwidth requirements are lower, less data is transmitted and computational resources at the server are saved.
Thus, this approach scales better for large networks.
Further, anomalous network traffic can be detected everywhere in the network.
This increases network coverage and robustness of the \ac{IDS}, as there is no more single point of failure.
Lastly, the \ac{IDS} can respond to malicious traffic quicker,
because the decision is made on the edge device itself and has no network delay on top. 
Furthermore, multiple \acp{IDS} could operate concurrently in a single network and exchange information.

Currently, there is a lack of distributed network-based \acp{IDS}, which can both run on low-power edge devices and account for the special networking requirements of \acp{ICS}.

To overcome this gap, in this paper we present a distributed \ac{IDS} tailored for industrial and sensor applications, using a statistical approach suitable for embedded low performance \acp{MCU}.
Compared to signature and rule-based approaches, this approach has some advantages, such as a dynamic learning without fixed rules, as well as the lack of periodic signatures updates of malicious software.
To demonstrate the ease of use, we implemented the \ac{IDS} on commodity hardware and validate its performance within a physical process.

Our work answers the following research questions about distributed network-based \acp{IDS} in industrial and sensor environments. 
Special concerns for distributed intrusion detection are:
\begin{itemize}
  \item How long does an \ac{ICS} network have to be observed 
        to get a sufficient amount of networking data to 
        \enquote{learn} its regular behavior?
  \item How to handle user interaction, e.g. through a \ac{HMI} 
        with its influence on system timings?
  \item How to retrieve the incident message from the edge node 
        and inform an operator about it? 
  \item What degree of deviation from the regular behavior of the network 
        is tolerable before detecting it as an incident?
\end{itemize}

Key contributions of our approach include:
\begin{itemize}
  \item We provide a easily portable implementation, 
        since only the widely used \acs{LwIP} stack with FreeRTOS is needed.
        Additionally, the porting of the approach is easy to handle.
  \item We provide a detailed performance analysis of an actual \ac{IDS} 
        implementation and evaluate the capabilities of the introduced approach
        in a realistic industrial control system environment.
\end{itemize}

The remainder of this paper is structured as follows:
\Cref{sec:background} provides necessary background knowledge.
\Cref{sec:methodology} explains the methodology for this approach. 
The \ac{PoC} implementation is described in \Cref{sec:implementation}.
Validation and benchmarking in an industrial test"~bed are done in \Cref{sec:measurement}.
Lastly, \Cref{sec:conclusion}\@ concludes this paper and gives an outlook.

\section{Background \label{sec:background}}
Industrial devices, networks and \acp{IDS} placed there have specific requirements because this is a different domain compared to the office environment.
A brief introduction, the state of the art and related work is summarized in the following.

\subsection{Industrial Control Systems and Communications\label{subsec:icscomms}} 
Traditionally, \acp{ICS} have operated in isolated, partially air"~gapped networks. 
As a result, IT"~security has not been a concern~\cite{igure2006security}. 
When interconnecting industrial devices, however, this is no longer acceptable. 
To make things worse, the industrial sector is dominated by proprietary and decades-old legacy devices and protocols, which offer little to no security on their own. 

Adding security to those devices and protocols is challenging, as industrial networks are often subject to real"~time requirements. 
To be able to meet these requirements, which include well-defined communication timings,
especially low performance components need all their power for the task at hand. 
As reliability, availability and predictable timings are paramount goals,
security mechanisms which introduce a large overhead are not acceptable.

Additionally, due to the highly heterogeneous nature of the network components in industrial environments and the lack of interoperability standards, enabling secure, concise and cost efficient communication between all network devices
remains a big challenge. 

Cardenas et al. already addressed different security problems of \acp{ICS} in 2008~\cite{cardenas2008research}.
However, in the paper by Cardenas et al., there are two example given at the end.
The first is the control of the process under \ac{DoS} attacks and the second is the detection of attacks.
Modern \acp{PLC} are
still vulnerable to, among others, simple \ac{DoS} attacks~\cite{niedermaier2018you}.

\subsection{Intrusion Detection in \acs{SCADA} Networks\label{sec:intrusiondetecton}}
A \ac{SCADA} system is a special \acs{ICS}, where the monitoring and controlling of technical processes is done by a computer system.
There is a large body of research in \acp{IDS} in \ac{ICS} and \ac{SCADA} networks. 
Works can roughly be divided in detecting compromise of networked \textit{devices}, e.g.~caused by malware or control-flow anomalies \cite{Reeves2012, Jin2018},
and the second group, which is concerned with detecting intrusive network \textit{traffic}, to which our work belongs to.

The first criterion is the \textit{type of data} used for classifying traffic into benign and intrusive.
For example, Liu and Liu show, how voltage drops reveal the presence of an attacker in RS485 daisy chain networks \cite{Liu2018}.
The simplicity of their approach comes at the cost of being tailored towards a particular protocol and requiring modeling of the network beforehand.
This illustrates one of the key problems for designing a generalized \ac{IDS} for \ac{SCADA} networks: A large number of networking protocols is encountered in the field.
However, to be able to properly scan the network,
a monitoring system must accurately dissect the traffic.
Thus, it requires a precise model of every single protocol it captures,
most of which are proprietary \cite{goldenberg2013accurate}.
In contrast, our system uses a metadata"~based approach which operates independently of the underlying transport protocol.
This overcomes the need of modeling the protocol.
Further, our approach brings flexibility and permits our system to be retrofitted to already deployed networks.

Network traffic can be acquired and processed either centralized or decentralized.
As stated in the introduction, the latter is preferable.
Additionally, passive acquisition has the benefit of not interfering with the network traffic and thus avoiding the risk to interrupt production processes.

While the strict timing requirements in \ac{ICS} traffic hinder the introduction of some security mechanisms, they can be exploited for distinguishing between normal and intrusive network traffic.
Barbosa et al.~utilize periodic cycle time to distinguish between normal and intrusive traffic \cite{barbosa2014anomaly}.
However, they do not implement a distributed method.
This must be considered insufficient with respect to an insider attacker, who can access the local network from anywhere within.
Lin et al.~also present a method which attempts timing"~based intrusion detection~\cite{lintiming}.
Yet again, as opposed to our works, their system captures data centrally. 
Further, their system uses network traffic capture files as an input with no real world testbed. 
In contrast, our system is deployed in a real"~life test"~bed and can adjust the baseline, which distinguishes between normal and intrusive traffic, during runtime.

Haller et al.~\cite{haller2019engineering} show the feasibility of an \ac{IDS} based on a monitoring task and the
statistical cumulative sum, running on a Phoenix Contact ILC 350-PN controller.
However, this system is a basic approach for
this specific \ac{PLC} and mostly only handles these two detection possibilities.

There is a lot of research going on in the field of intrusion detection.
Some of the published concepts are summarized
in a survey on \acp{IDS} in wireless networks~\cite{butun2014survey}.
These methods have also been used within industrial networks
with adjustments to its specific requirements.
Another survey on \acp{IDS} and \acp{IPS} in \ac{SCADA} networks has been published by Zhu et al.~\cite{zhu2010scada}.
However, in this surveys no distributed network analysis on low performance \acp{MCU} are handled.

Further work was done by Zimmer et
al.~\cite{zimmer2015intrusion}, who introduce security building blocks for real"~time \acp{CPS}.
These are based on measurements
of the real"~time operation system with no focus to the network data analysis.

Payer shows a state"~driven \ac{IDS}
implemented within the LwIP Stack in 2003~\cite{payer2003}.
It analyses the connection states of a connection and compares them with stored database.
However, this work covers only a few scenarios and does not perform well as a network \ac{IDS}.

The previous work mostly focuses on a purely network"~based approach with a high effort
necessary for integration, or host based systems with high requirements to 
the computing power of the host.
In contrast to previous work, we provide the following advantages:

\begin{itemize}
	\item Distributed \ac{IDS} on industrial edge node devices, e.g. sensors.
	      This does not require changes to the network infrastructure while listening to  network traffic.
	\item Analysis of the periodic occurrence of packets/requests.
	\item Approach and implementation are protocol"~independent, because they utilize meta information. 
	      This enables later usage with cryptographic protection mechanisms.
\end{itemize}

\section{Methodology \label{sec:methodology}}
This section sheds light on what the \ac{IDS} should protect against, 
which data can be used for the \ac{IDS} analysis and how the intrusion is announced to e.g.~the operator.
Further, we discuss the strengths and weaknesses of our approach.

\subsection{Attacker Model \label{subsec:attackermodel}}
In this paper, we consider both local and remote attackers.

An attacker with \textbf{local access} could, for example, be an employee or a visitor.
This attacker has both physical and network access.
Being physically present, the attacker is able to remove edge nodes from the network by e.g.~unplugging the Ethernet cable.
Furthermore, he can launch simple attacks such as pressing the emergency stop button at a machine, resulting in a denial of service attack.
From within the network perspective, the goal of the attacker is to be able to eavesdrop and manipulate messages from sensor nodes.
Also, he can inject arbitrary messages or delay messages in order to launch a denial of service attack.

The \textbf{remote access} attacker has gained access to the internal network over the Internet or can directly attack exposed devices.
Flooding attacks, which lead to a \ac{DoS}, as well as spoofing and injecting messages can be conducted.
\ac{MitM} attacks are not possible for the remote attacker.

\subsection{Overview on Workflow \label{subsec:workflow}}
The \ac{IDS} operates by firstly observing the network traffic in a non"~corrupted network, learning \textit{normal} or \textit{benign} traffic.
Utilizing the deterministic timings in traffic flows, it derives periodicity"~thresholds, which separate normal from intrusive traffic.
Calculating the thresholds is accomplished in two ways.
Firstly, the metadata of each connection are analyzed and categorized.
Secondly, these categorized connections are then analyzed based on their periodicity. 
After the learning phase, the traffic of the live \ac{ICS} is compared to these thresholds and classified as either normal or intrusive.
To account for slight changes in the \ac{ICS} traffic behavior, thresholds are adjusted during runtime.

\subsubsection{Meta Data Selection \label{sec:analyzeddata}}
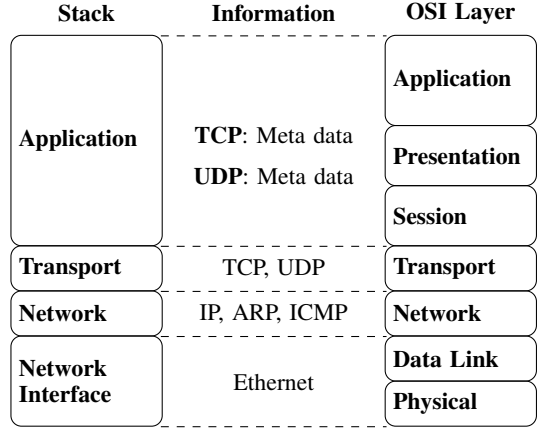
\begin{figure}[htb]
	\centering
	\begin{tikzpicture}[node distance=1cm,
	auto,
	block/.style={
		rectangle,
		draw=black,
		align=center,
		rounded corners,
		dashed
	}
	]
	\coordinate (a) at (0,0.8);
	\coordinate (b) at (0,1.4);
	\coordinate (c) at (0,2.0);
	\coordinate (d) at (0,2.6);
	\coordinate (e) at (0,3.2);
	\coordinate (f) at (0,4.0);
	\coordinate (g) at (0,4.8);
	\coordinate (h) at (0,6.0);
	
	\coordinate (i) at (7,0.8);
	\coordinate (j) at (7,1.4);
	\coordinate (k) at (7,2.0);
	\coordinate (l) at (7,2.6);
	\coordinate (m) at (7,3.2);
	\coordinate (n) at (7,4.0);
	\coordinate (o) at (7,4.8);
	\coordinate (p) at (7,6.0);

	\small
	\draw[-,dashed] ([xshift=2.0cm]a) -- ([xshift=-2.0cm]i);
	\draw[-,dashed] ([xshift=2.0cm]c) -- ([xshift=-2.0cm]k);
	\draw[-,dashed] ([xshift=2.0cm]d) -- ([xshift=-2.0cm]l);
	\draw[-,dashed] ([xshift=2.0cm]e) -- ([xshift=-2.0cm]m);
	\draw[-,dashed] ([xshift=2.0cm]h) -- ([xshift=-2.0cm]p);
	
	\node[block, draw, align=left, minimum width=2cm, text width=1.8cm, solid, minimum height=1.2cm, anchor=south west] at (a)   {\textbf{Network Interface}};
	\node[block, draw, align=left, minimum width=2cm, text width=1.8cm, solid, minimum height=0.6cm, anchor=south west] at (c)   {\textbf{Network}};
	\node[block, draw, align=left, minimum width=2cm, text width=1.8cm, solid, minimum height=0.6cm, anchor=south west] at (d)   {\textbf{Transport}};
	\node[block, draw, align=left, minimum width=2cm, text width=1.8cm, solid, minimum height=2.8cm, anchor=south west] at (e)   {\textbf{Application}};
	
	\node[block, draw, align=left, minimum width=2cm, text width=1.8cm, solid, minimum height=0.6cm, anchor=south east] at (i)   {\textbf{Physical}};
	\node[block, draw, align=left, minimum width=2cm, text width=1.8cm, solid, minimum height=0.6cm, anchor=south east] at (j)   {\textbf{Data Link}};
	\node[block, draw, align=left, minimum width=2cm, text width=1.8cm, solid, minimum height=0.6cm, anchor=south east] at (k)   {\textbf{Network}};
	\node[block, draw, align=left, minimum width=2cm, text width=1.8cm, solid, minimum height=0.6cm, anchor=south east] at (l)   {\textbf{Transport}};
	\node[block, draw, align=left, minimum width=2cm, text width=1.8cm, solid, minimum height=0.8cm, anchor=south east] at (m)   {\textbf{Session}};
	\node[block, draw, align=left, minimum width=2cm, text width=1.8cm, solid, minimum height=0.8cm, anchor=south east] at (n)   {\textbf{Presentation}};
	\node[block, draw, align=left, minimum width=2cm, text width=1.8cm, solid, minimum height=1.2cm, anchor=south east] at (o)   {\textbf{Application}};
	
	\node[align=center, minimum width=2.9cm, text width=2.8cm, minimum height=1.6cm, anchor=center] at ($(a)!0.5!(k)$)   {\small Ethernet};
	\node[align=center, minimum width=2.9cm, text width=2.8cm, minimum height=0.8cm, anchor=center] at ($(c)!0.5!(l)$)   {\small IP, ARP, ICMP};
	\node[align=center, minimum width=2.9cm, text width=2.8cm, minimum height=0.8cm, anchor=center] at ($(d)!0.5!(m)$)   {\small TCP, UDP};
	\node[align=center, minimum width=2.9cm, text width=2.8cm, minimum height=2.4cm, anchor=south] at ($(e)!0.5!(m)$)   {\small \textbf{TCP}: Meta data \\[0.2cm] \textbf{UDP}: Meta data};	
	\node[align=center, minimum width=1.9cm, text width=1.8cm, minimum height=0.6cm, anchor=south west] at (h)   {\textbf{Stack}};
	\node[align=center, minimum width=1.9cm, text width=2.8cm, minimum height=0.6cm, anchor=south] at ($(h)!0.5!(p)$)   {\textbf{Information}};
	\node[align=center, minimum width=1.9cm, text width=1.8cm, minimum height=0.6cm, anchor=south east] at (p)   {\textbf{OSI Layer}};
	\normalsize
	\end{tikzpicture}
	\vspace*{0.2mm}
	\caption{Used information from the stack for intrusion detection.}
	\label{fig_protocols}
\end{figure}

First, we discuss, which features of the network traffic are suitable to be used for characterizing and classifying the connections.
Basically the introduced \ac{IDS} approach can analyze, train and then make decisions using \textit{all} network data it receives.
\Cref{fig_protocols} shows the meta information, which can be analyzed by the \ac{IDS}, mapped to the layers of the network stack according to the \ac{OSI} model.

Analyzed metadata is chosen such, that the \acp{IDS} operates completely protocol"~independent.
In this case, metadata is all data, which resides below the application layer.
Thus, no further information is needed during the deployment, which means, that the operator does not have to set rules.
Specifically, the following information is used:
\begin{itemize}
	\item Source and destination ports are used from the \textbf{\ac{TCP}} 
	      and \textbf{\ac{UDP}} header.
	\item From the \textbf{\ac{IP}} header the source address 
	      and destination address are used.
	\item \textbf{\ac{ARP}} requests and responses contain \ac{MAC} 
	      and \ac{IP} addresses, which are mapped to each other.
	\item The \textbf{Ethernet} header contains the destination \ac{MAC} address 
	      and the source \ac{MAC} address.
	      Those must be consistent with each other in order to recognize
	      e.g.~\ac{ARP} poisoning.
	\item Optionally, \textbf{meta data} from the application layer could be used.
\end{itemize}

The source and destination ports are typically unfeasible for intrusion detection because they they vary:
As soon as a connection is terminated, usually a new port is used by the client.
This must be taken into account when categorizing the connections.
It is still feasible to use the port on the server side, e.g.~port 502 for Modbus/TCP to analyze the network packets.
The other metadata, however, remain consistent and does not change at reconnections.
Allocating metadata from connections is already a good starting point for intrusion detection.
This is similar to rule-based detection, which is also used in e.g.~firewalls.

\subsubsection{Exploiting and Learning the Network Timing Behavior \label{subsec:learning}}
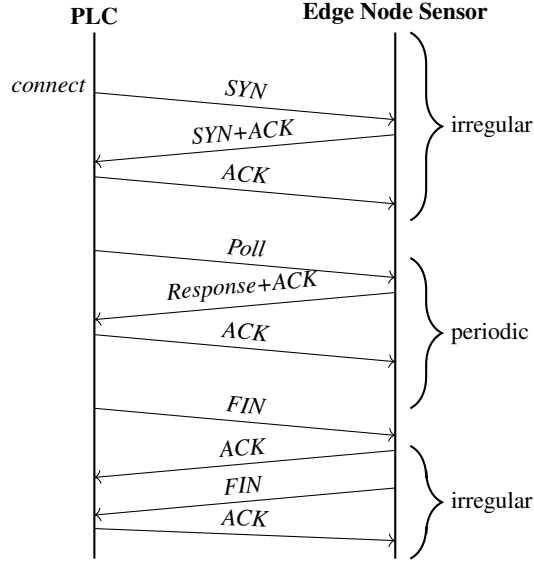
\begin{figure}[htb]
	\centering
	\begin{footnotesize}
		\begin{tikzpicture}[node distance=1cm,
		auto
		]
		\coordinate (A) at (2,7);
		\coordinate (B) at (2,0);
		\coordinate (C) at (6,7);
		\coordinate (D) at (6,0);
		\draw[thick] (A)--(B) (C)--(D);
		\draw (A) node[above]{\textbf{PLC} };
		\draw (C) node[above]{\textbf{Edge Node Sensor}};
		
		\coordinate (E) at ($(A)!0.1!(B)$);
		\draw (E) node[left]{\textit{connect}};
		
		\coordinate (F) at ($(C)!0.18!(D)$);
		\draw (F) node[right]{\textit{}};
		\draw[->] ([xshift=0.0cm, yshift=-0.1cm]E) -- ([xshift=0.0cm, yshift=0.1cm]F) node[midway,sloped,above]{\textit{\small SYN}};
		
		\coordinate (G) at ($(A)!0.26!(B)$);
		\draw (G) node[left]{\textit{}};
		\draw[->] ([xshift=0.0cm, yshift=-0.1cm]F) -- ([xshift=0.0cm, yshift=0.1cm]G) node[midway,sloped,above]{\textit{\small SYN+ACK}};
		
		\coordinate (H) at ($(C)!0.34!(D)$);
		\draw (H) node[right]{\textit{}};
		\draw[->] ([xshift=0.0cm, yshift=-0.1cm]G) -- ([xshift=0.0cm, yshift=0.1cm]H) node[midway,sloped,above]{\textit{\small ACK}};
		
		\draw[thick,black,decorate,decoration={brace,amplitude=12pt,mirror}] ([xshift=0.2cm, yshift=-2.5cm]C) -- ([xshift=0.2cm, yshift=0.0cm]C) node[midway, right,xshift=12pt,]{irregular};
		
		\coordinate (I) at ($(A)!0.40!(B)$);
		\draw (I) node[left]{\textit{}};
		
		\coordinate (J) at ($(C)!0.48!(D)$);
		\draw (J) node[right]{\textit{}};
		\draw[->] ([xshift=0.0cm, yshift=-0.1cm]I) -- ([xshift=0.0cm, yshift=0.1cm]J) node[midway,sloped,above]{\textit{\small Poll}};
		
		\coordinate (K) at ($(A)!0.56!(B)$);
		\draw (K) node[left]{\textit{}};
		\draw[->] ([xshift=0.0cm, yshift=-0.1cm]J) -- ([xshift=0.0cm, yshift=0.1cm]K) node[midway,sloped,above]{\textit{\small Response+ACK}};
		
		\coordinate (L) at ($(C)!0.64!(D)$);
		\draw (L) node[right]{\textit{}};
		\draw[->] ([xshift=0.0cm, yshift=-0.1cm]K) -- ([xshift=0.0cm, yshift=0.1cm]L) node[midway,sloped,above]{\textit{\small ACK}};
		
		\draw[thick,black,decorate,decoration={brace,amplitude=12pt,mirror}] ([xshift=0.2cm, yshift=-5.0cm]C) -- ([xshift=0.2cm, yshift=-3.0cm]C) node[midway, right,xshift=12pt,]{periodic};
		
		\coordinate (M) at ($(A)!0.70!(B)$);
		\draw (M) node[left]{\textit{}};
		
		\coordinate (N) at ($(C)!0.78!(D)$);
		\draw (N) node[right]{\textit{}};
		\draw[->] ([xshift=0.0cm, yshift=-0.1cm]M) -- ([xshift=0.0cm, yshift=0.1cm]N) node[midway,sloped,above]{\textit{\small FIN}};
		
		\coordinate (O) at ($(A)!0.86!(B)$);
		\draw (O) node[left]{\textit{}};
		\draw[->] ([xshift=0.0cm, yshift=-0.1cm]N) -- ([xshift=0.0cm, yshift=0.1cm]O) node[midway,sloped,above]{\textit{\small ACK}};
		\draw[->] ([xshift=0.0cm, yshift=-0.6cm]N) -- ([xshift=0.0cm, yshift=-0.4cm]O) node[midway,sloped,above]{\textit{\small FIN}};
		\draw[->] ([xshift=0.0cm, yshift=-1.7cm]M) -- ([xshift=0.0cm, yshift=-1.3cm]N) node[midway,sloped,above]{\textit{\small ACK}};
		
		\draw[thick,black,decorate,decoration={brace,amplitude=12pt,mirror}] ([xshift=0.2cm, yshift=-7.0cm]C) -- ([xshift=0.2cm, yshift=-5.5cm]C) node[midway, right,xshift=12pt,]{irregular};
		\end{tikzpicture}
	\end{footnotesize}
	\caption{Timing in sensor networks with polling, separated in periodic and irregular timings.}
	\label{fig:timing}
\end{figure}

A defining characteristic of an \ac{ICS} network is periodical \textit{polling} of inputs and outputs. 
This creates a homogeneous timing picture of the connections in the network.
In the example in \Cref{fig:timing}, first a TCP/IP connection is established~(SYN).
The timing pattern here is considered irregular.
The same holds true for closing the connection~(FIN).

However, after having established the connection, in the center of \Cref{fig:timing} we can observe the periodic timings caused by periodic polling.

Periodic behavior is exploited in the proposed approach and observed in the initial training phase.
During this training phase, we require the network to be untainted.

We use two methods to derive thresholds, which separate normal behavior from an intrusion.
By filtering one specific connection, e.g.~from one edge node to the central \ac{PLC}. the time series is getting periodic.
This mostly depends on the network infrastructure and implementation.
Nevertheless, this only requires a longer learning time, if the traffic is more irregular.

\textbf{Statistical Analysis of Normal Traffic Behavior}:
During the learning phase, the minimum and maximum interarrival time is calculated.
The values $t_l$ are recorded during the learning phase, which is the interarrival time between packets of the same connection type.

The \ac{IDS} forms a cumulative moving average (see \Cref{eq:idsaverage}) over the interarrival time of the packets.
The interarrival time during learning is $t_l$ and in active mode $t$.
To calculate a specific mean value, the amount of used interarrival times is divided by $n_l$ during learning respectively $n$ in active mode. 
This is used after the learning phase as a reference to detect changes in the frequency of packet transfer in the specific connections.
Since there may still be minimal deviations after the learning phase, an offset ($\Delta$) is added on top.

\begin{equation} \label{eq:idsaverage}
\footnotesize
\frac{t_{l1} + ... + t_{ln}}{n_l}*(1-\Delta) < \frac{t_{1} + ... + t_{n}}{n} < \frac{t_{l1} + ... + t_{ln}}{n_l}*(1+\Delta)
\end{equation}

After the learning phase, the moving average is further calculated and compared with the trusted reference.
If this current moving average is outside of this trusted reference, an anomaly is assumed.
Those calculations must be done for each network connection to the edge node device.

\textbf{Minimum and Maximum Classification of Traffic}:
After the \ac{IDS} is switched from the learning phase to active, the current packet interarrival time ($t_{current}$) is compared with the previous calculated maximum and minimum.
If this is outside of those boundaries, this is seen as an intrusion (see \Cref{eq:idscompare}).
In addition, an adjustable offset ($\Delta$) is also specified for the minimum and maximum limits. 

\begin{equation} \label{eq:idscompare}
\footnotesize
min(t_{l1}, t_{l2}, ..., t_{ln})*(1-\Delta) < t_{current} < max(t_{l1}, t_{l2}, ..., t_{ln})*(1+\Delta)
\end{equation}

Network traffic is categorized and statistically evaluated using the parameters described in \Cref{sec:analyzeddata}.
There are parameters like \ac{IP} and \ac{MAC} address, which must not deviate after the learning phase. 
In contrast, there are metadata such as the time behavior, 
which accept a certain tolerance to avoid false reports.
On the one hand, a recognized intrusion should be processed and displayed directly at the edge node, as well as transmitted to a central logging server.

\subsection{Intrusion Announcement \label{sec:intrusionannouncement}}
We use broadcasts to announce an intrusion to the network.
As a result, the central intrusion logger only needs to be in the same broadcast domain and no configuration is necessary.
Additionally, in order to detect \ac{DoS} attacks which are supposed to block the messages, a keep alive message is sent at a certain time interval.
If the central logger does not receive this message within a predetermined timeout interval, an intrusion is assumed to take place.
Furthermore, the message must contain a signature and a changing variable to prevent replay attacks.
In addition to the centralized logging, local signaling can be used to warn operators within the plant.

\subsection{Discussion of Strenghts and Limitations \label{sec:requirements}}
One of the biggest strengths of this approach is the easy integration into existing networks,
because no special network hardware such as as mirror ports are necessary.
In addition, each edge node device can defend itself and does not have to trust other parties.

In order to get a trusted comparison base, it has to be ensured that there is no attacker in the network during the initial learning phase.
To launch an attack undetectable to the \ac{IDS}, the attacker has to generate the same traffic used in the training phase, otherwise it will be recognized as an anomaly.

\section{Implementation \label{sec:implementation}}
In this section, we demonstrate the feasibility of our distributed \ac{IDS} approach by providing an implementation on real low-performance \acp{MCU}.
This sets our work apart from many other proposed \acp{IDS}, which only use simulation for validation.

\subsection{Hardware}
\begin{figure}[htb]
\centering
    \begin{tikzpicture}    
    \node[inner sep=0pt] (shield) at (0,0)
            {\includegraphics[width=0.90\columnwidth]{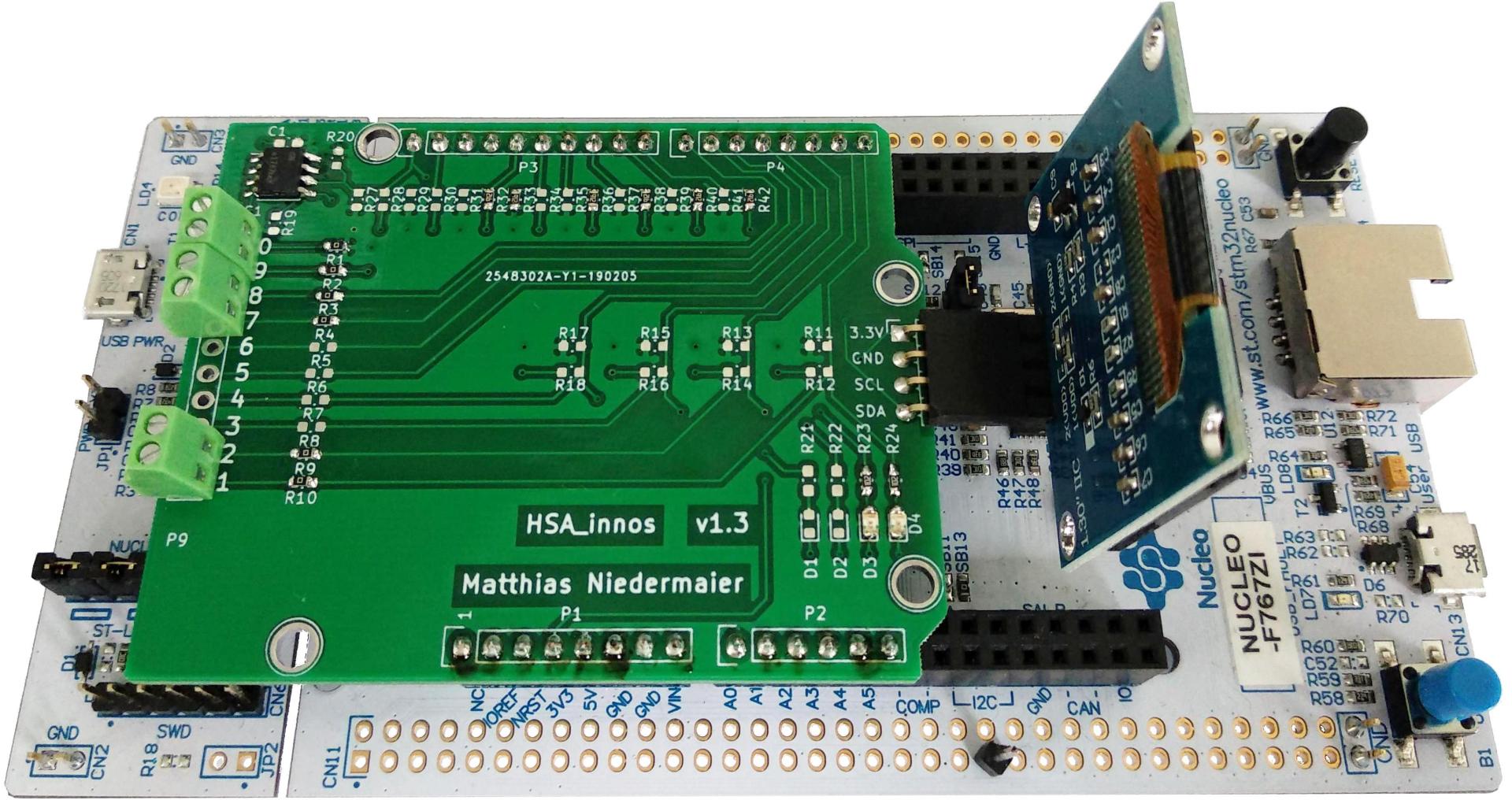}};
        \node[align=left,anchor=east] at (1.0,3.3) {Display};
        \draw[->,color=red!50, line width=0.5mm] (1.0,3.3) -- (3.0,0.7);
        \node[align=left,anchor=east] at (3.8,4.0) {Ethernet};
        \draw[->,color=red!50, line width=0.5mm] (3.8,4.0) -- (5.7,0.8);
        \node[align=left,anchor=west] at (-3.5,3.5) {Inputs};
        \draw[->,color=red!50, line width=0.5mm] (-3.5,3.5) -- (-5.0,1.3);
        \node[align=left,anchor=west] at (-3.0,3.0) {Outputs};
        \draw[->,color=red!50, line width=0.5mm] (-3.0,3.0) -- (-5.2,-0.6);
        \node[align=left,anchor=west] at (-3.5,-4.6) {NUCLEO-F767ZI Development Board};
        \draw[->,color=red!50, line width=0.5mm] (0.0,-4.4) -- (0.0,-3.6);
        \node[align=left,anchor=west] at (3.5,-4.4) {LEDs};
        \draw[->,color=red!50, line width=0.5mm] (3.5,-4.4) -- (0.9,-1.1);
    \end{tikzpicture}
	\caption{Picture of the baseboard and the custom PCB of the used edge node device.}
	\label{fig:edgepcb}
\end{figure}
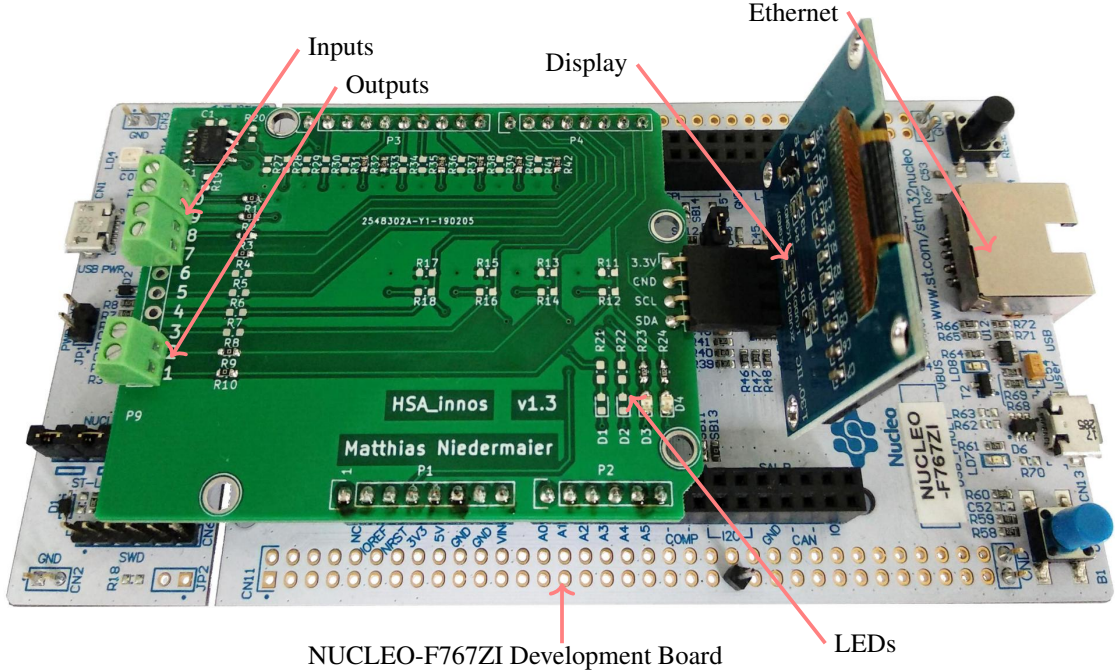

The edge node devices are implemented on ST NUCLEO-F767ZI\footnote{\url{https://www.st.com/en/evaluation-tools/nucleo-f767zi.html}} development boards.
They have an ARM Cortex-M7 core, which is operating at a frequency of 216 MHz with 512 kB of RAM and 2 MB flash.
The board is equipped with an Ethernet transceiver and corresponding RJ45 jack.
\Cref{fig:edgepcb} shows the used development board from STM and the custom designed \ac{PCB} op top.
This is used to control the \acp{IO} and the display, which is controlled over \ac{I2C}.

\subsection{Stack integration}
\begin{figure}[htb]
\centering
\begin{tikzpicture}[node distance=1cm,
    auto,
    block/.style={
      rectangle,
      draw=black,
      align=center,
      rounded corners,
      dashed
    }
  ]
  \coordinate (a) at (1,0.0);
  \coordinate (b) at (1,1.0);
  \coordinate (c) at (1,2.0);
  \coordinate (d) at (1,3.2);
  
  \node[rectangle, draw, align=left, minimum width=6.0cm, text width=6cm,
      minimum height=0.5cm, anchor=south west] at (a.south west) (l) {Device Driver};  
  \node[rectangle, draw, align=left, minimum width=5.0cm, text width=5cm,
      minimum height=0.5cm, anchor=south west] at (b.south west) (m) {Middleware: FreeRTOS, LwIP, ...};  
  \node[rectangle, draw, align=left, minimum width=4.0cm, text width=4cm,
      minimum height=0.7cm, anchor=south west] at (c.south west) (n) {LwIP API};  
  \node[rectangle, draw, align=center, minimum width=1.0cm, text width=1cm, color=black!20!red,
      minimum height=0.5cm, anchor=south west] at ([xshift=2.5cm, yshift=0.1cm]c.south west) (x)
      {\small \textcolor{black!20!red}{IDS}};  
  \node[rectangle, draw, align=left, minimum width=6.0cm, text width=6cm,
      minimum height=0.7cm, anchor=south west] at (d.south west) (o) {Application};  
  \node[rectangle, draw, align=center, minimum width=3.4cm, text width=3.3cm, color=black!50!white,
      minimum height=0.5cm, anchor=south west] at ([xshift=1.95cm, yshift=0.1cm]o.south west) (x)
      {\small \textcolor{black!50!white}{Modbus/TCP, WEB, ...}}; 
      
  \draw [-{Stealth[scale=1.0]}] 
      ([xshift=0.5cm, yshift=0.0cm]l.north) to
      ([xshift=1.0cm, yshift=0.0cm]m.south);
  \draw [{Stealth[scale=1.0]}-] 
      ([xshift=-1.0cm, yshift=0.0cm]l.north) to
      ([xshift=-0.5cm, yshift=0.0cm]m.south);
  \draw [-{Stealth[scale=1.0]}] 
      ([xshift=0.5cm, yshift=0.0cm]m.north) to
      ([xshift=1.0cm, yshift=0.0cm]n.south);
  \draw [{Stealth[scale=1.0]}-] 
      ([xshift=-1.0cm, yshift=0.0cm]m.north) to
      ([xshift=-0.5cm, yshift=0.0cm]n.south);
  \draw [-{Stealth[scale=1.0]}] 
      ([xshift=1.5cm, yshift=0.0cm]n.north) to
      ([xshift=0.5cm, yshift=0.0cm]o.south);
  \draw [{Stealth[scale=1.0]}-] 
      ([xshift=-0.0cm, yshift=0.0cm]n.north) to
      ([xshift=-1.0cm, yshift=0.0cm]o.south);
  \draw [-{Stealth[scale=1.0]}] 
      ([xshift=2.5cm, yshift=0.0cm]m.north) to
      ([xshift=2.0cm, yshift=0.0cm]o.south);
  \draw [{Stealth[scale=1.0]}-] 
      ([xshift=2.0cm, yshift=0.0cm]m.north) to
      ([xshift=1.5cm, yshift=0.0cm]o.south);
  \draw [-{Stealth[scale=1.0]}] 
      ([xshift=2.5cm, yshift=0.0cm]l.north) to
      ([xshift=2.5cm, yshift=0.0cm]o.south);
  \draw [{Stealth[scale=1.0]}-] 
      ([xshift=3.0cm, yshift=0.0cm]l.north) to
      ([xshift=3.0cm, yshift=0.0cm]o.south);
  \end{tikzpicture}
\caption{Overview of the system with integration of the \ac{IDS} into LwIP.}
\label{fig:idslwip}
\end{figure}
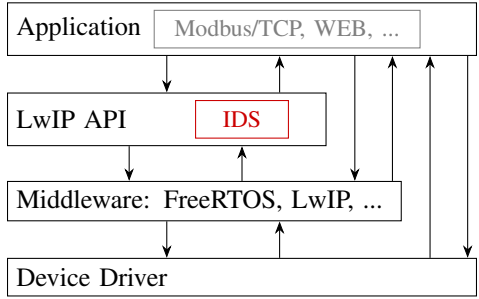

As shown in \Cref{fig:idslwip}, the \ac{IDS} is integrated into the widely used LwIP \cite{dunkels2001design} stack.
All RX and TX data is processed by the \ac{IDS} before they are forwarded to the regular \ac{API} of the \ac{LwIP} stack. 
This design is highly beneficial, as the \ac{IDS} can be used without changes to existing projects, which use the \ac{LwIP} \ac{API}.

The device driver is the \ac{HAL} between the operating system and the hardware.
It permits easy access to the hardware and enables replacing the \acp{MCU} with low effort.
This \ac{HAL} is used by FreeRTOS, the LwIP stack and applications, which need direct hardware access.
The \ac{LwIP} stack running as a middleware provide the \ac{API} for receiving and sending packets, which is used to integrate our \ac{IDS}.
This is used by the applications, e.g.~the Modbus/TCP server to interact within an industrial environment.
The Modbus/TCP and the web server are self implemented light weight applications.

Another advantage of using the existing \ac{LwIP} \ac{API} is, that the \ac{IDS} can also be used as a \ac{IPS}. In this case, packets, which are detected as an intrusion, are not passed to the application layer and discarded instead.

\subsection{Centralized Logging of Intrusions}
In our decentralized \ac{IDS}, every sensor node individually captures and preprocesses network data.
However, to permit technicians centralized administration of edge nodes, a central logging is necessary.
Every sensor is reporting the current ``security'' state in a periodic way to this central logger.
If the sensor does not send the status message within a certain time frame, 
because e.g.~an attacker is flooding the network,
the monitoring and logging server must detect and report this as an incident.
The key features of the centralized intrusion and alive notification are:

\begin{itemize}
\item The intrusion message is send to the network via \ac{UDP} broadcast.
\item The message is signed with an \ac{HMAC}"~based using a \ac{PSK}.
\item The message is send out every 10 seconds, with status information and keep alive message.
\item The system time is part of the message for replay protection.
\end{itemize}

These messages are gathered by the logging server and if there is no keep alive message within 20 seconds, the host is regarded as contaminated.


\subsection{Intrusion Notification on the Edge Node}
\begin{figure}[htb]
	\centering
	    \begin{tikzpicture}
	\node[inner sep=0pt] (shield) at (0,0)
            {\includegraphics[width=0.35\columnwidth]{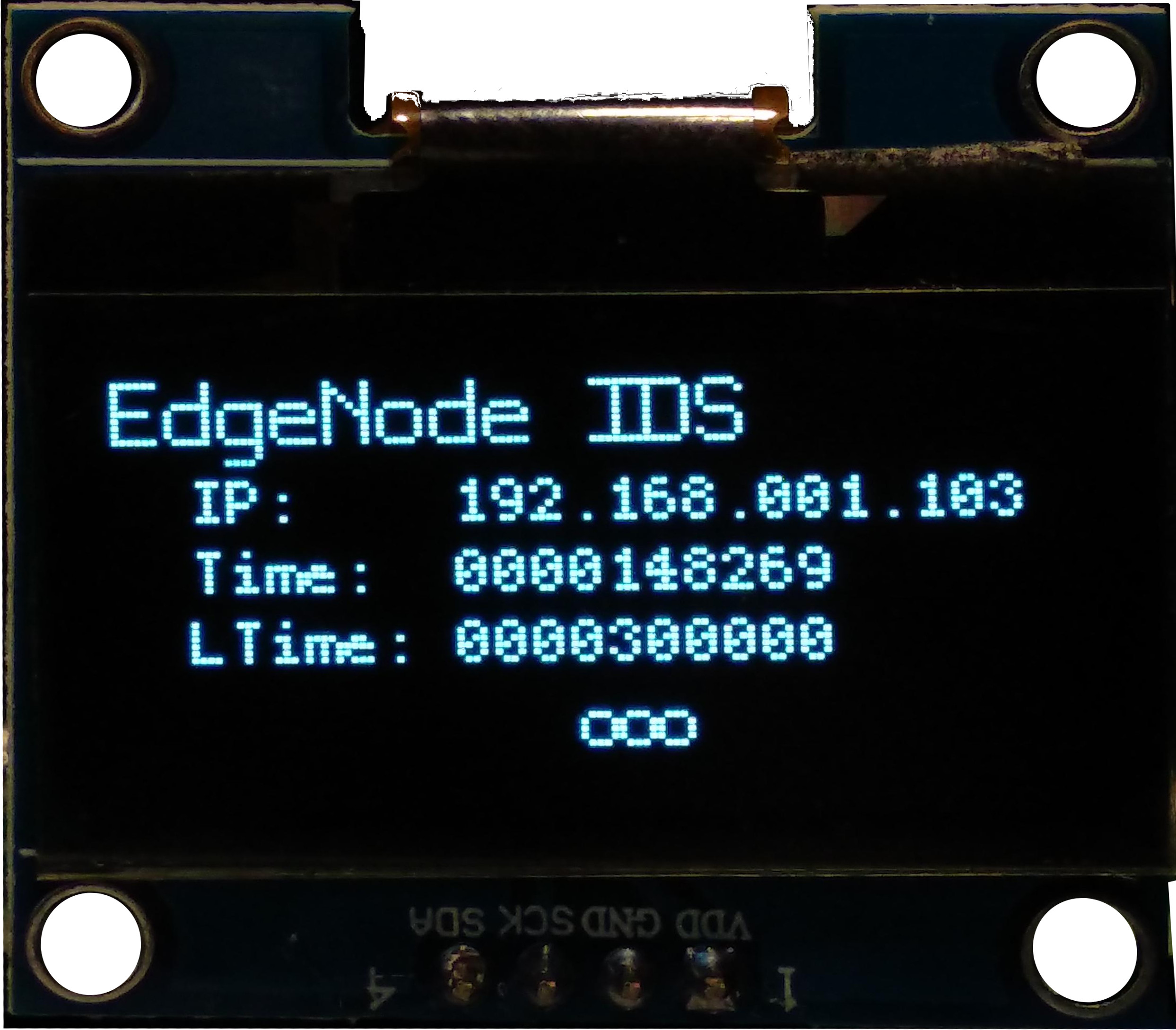}};
    \end{tikzpicture}
	\caption{Current state of the IDS on the edge node display.}
	\label{fig:ids_display}
\end{figure}

In addition, to the centralized intrusion logging, each edge node device can visualize its intrusion status using a display and/or a \ac{LED}.
This helps technicians in the control room, who have been warned of an intrusion, to quickly localize the affected edge nodes in the field.
Without information at the devices or dedicated tools such as \textit{EyeSec}, this has been proven to be a tedious task \cite{StriegelEtAl2019}.
\Cref{fig:ids_display} shows the current status of the edge node device.

After network startup, the current system time and the configured learning time is displayed.
If an intrusion occurs, this will be displayed and the red \ac{LED} on the baseboard lights up.
All these features are additional tools, which aid the operators in managing network security.

\section{Evaluation and Measurement Results \label{sec:measurement}}
To show the performance of the proposed approach, the edge node \ac{IDS} is evaluated in an open source industrial testbed.
Additionally, measurements were carried out and the detection scenarios were identified. 

\subsection{Evaluation in an Open Source Testbed}
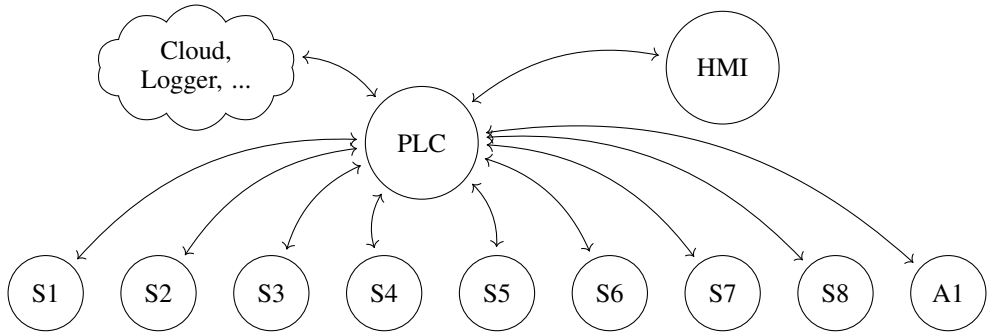
\begin{figure}[htb]
  \centering
\begin{tikzpicture}[node distance=1cm, 
    shorten >= 3pt,shorten <= 3pt,
    auto]    
    
  \foreach \x in {0,...,7}
    {\pgfmathtruncatemacro{\label}{\x +1}
    \node [circle, draw, fill=white, text width=0.5cm, minimum size=1.0cm,
      inner ysep=0.1cm, text centered] at (\x*1.5,0) (n\label) {S\label};}   
   
  \node [circle, draw, fill=white, text width=0.5cm, minimum size=1.0cm,
    inner ysep=0.1cm, text centered] at (12,0) (n9) {A1};
    
  \node [circle, draw, fill=white, text width=1.0cm, minimum size=1.5cm,
    inner ysep=0.1cm, text centered] at (5,2) (plc) {PLC};
    
  \node [circle, draw, fill=white, text width=1.0cm, minimum size=1.5cm,
    inner ysep=0.1cm, text centered] at (9,3) (hmi) {HMI};

  \node [cloud, draw, fill=white,cloud puffs=10,cloud puff arc=120, aspect=2, 
    text width=1.5cm,
    inner ysep=0.1cm, text centered] at (2,3) (c1) {Cloud, Logger, ...};
  
  \foreach \x in {1,...,4}
    \draw [<->, bend angle=25, bend left] (n\x) 
      to node[below, text width=1.5cm, text centered]{} (plc);
  \foreach \x in {5,...,9}
    \draw [<->, bend angle=25, bend right] (n\x) 
      to node[below, text width=1.5cm, text centered]{} (plc);
      
  \draw [<->, bend angle=25, bend left] (c1) 
      to node[below, text width=1.5cm, text centered]{} (plc);
  \draw [<->, bend angle=25, bend right] (hmi) 
      to node[below, text width=1.5cm, text centered]{} (plc);
     \end{tikzpicture}
\caption{Network system view on a ''standard'' industrial network mapped to our PoC test-bed implementation. Eight ''intelligent'' edge node sensors, one ''intelligent'' actuator, a PLC, a HMI and possibilities for cloud services.}
\label{fig:networksetup}
\end{figure}

Our open source testbed used in the evaluation is based on current \ac{ICS} operating architectures.
\Cref{fig:networksetup} shows such a common \ac{ICS} architecture, where the \ac{PLC} serves as a central data hub.
Please note, that our approach does not depend on a particular network architecture.
We only require periodic network traffic patterns.

The architecture used in this open source testbed contains 8 sensors (S1-S8), one actuator (A1), a \acs{PLC}, an \acs{HMI} and a connection to other services like \acs{SCADA}, logging and cloud applications.
Modbus/TCP is used as the industrial communication protocol, because it is widely used~\cite{swales1999open}.
Additionaly, Modbus/TCP offers manifold attack paths~\cite{huitsing2008attack}.
Thus, the \ac{IDS} can be evaluated and benchmarked in multiple realistic attack scenarios.

\Cref{tab:devices} summarizes the components used in the open source testbed.
In contrast to existing testbeds, software can be changed easily.
Further, as standard components are be used for evaluation, this testbed can be realized at low cost
~\cite{niedermaier2018cort}.
In contrast, with standard Modbus/TCP sensors, it is usually not possible to make changes such as inserting an \ac{IDS}.
Furthermore, measurements on the system are therefore not feasible, because no measurement routines could be used within the software on the device.

\begin{table}[htb]
    \centering
    \caption{Overview of devices used in the testbed.} 
    \label{tab:devices}
        \begin{tabular}{l l l l l}
            \hline
            \textbf{Identifier} & \textbf{Device} & \textbf{Software} & \textbf{Hardware} & \textbf{IP}\\
            \hline
            S1-S8               & Sensor          & FreeRRTOS, LwIP       & STM32F7           & 192.168.1.101-108\\
            A1                  & Actor           & FreeRRTOS, LwIP       & STM32F7           & 192.168.1.109\\
            PLC                 & PLC             & OpenPLCv3             & Raspberry Pi      & 192.168.1.50\\
            HMI                 & HMI             & Custom                & Raspberry Pi      & 192.168.1.40\\
            Cloud               & Cloud           & ScadaBR, Logging, ... & APU2C4            & 192.168.1.1\\
        \end{tabular}
\end{table}

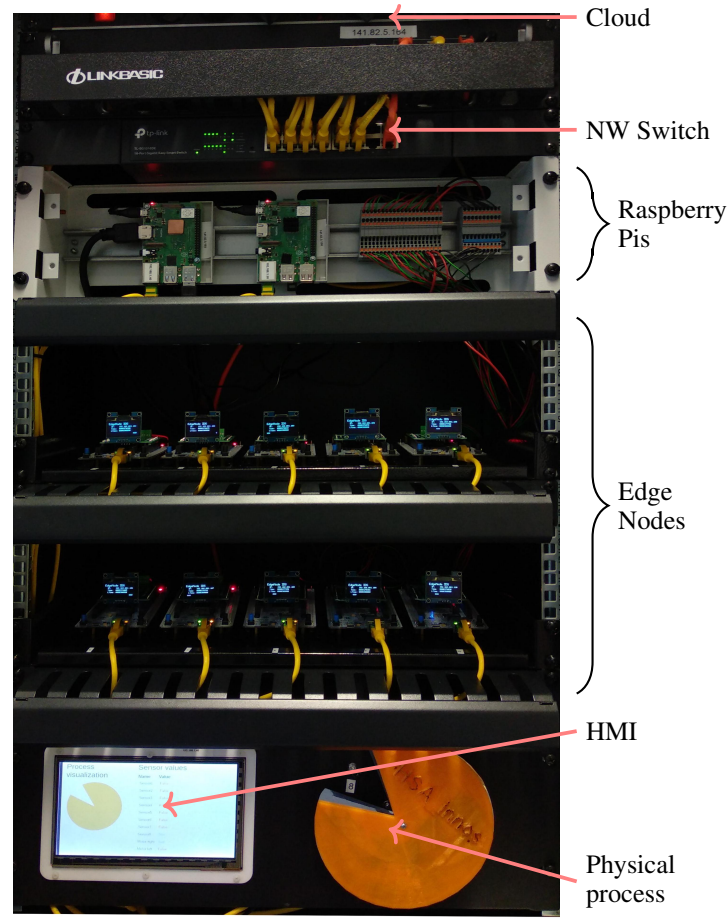
\begin{figure}[htb]
\center
\begin{tikzpicture}
\node[inner sep=0pt, anchor=south west] (x) at (0,0)
    {\includegraphics[width=.44\columnwidth]{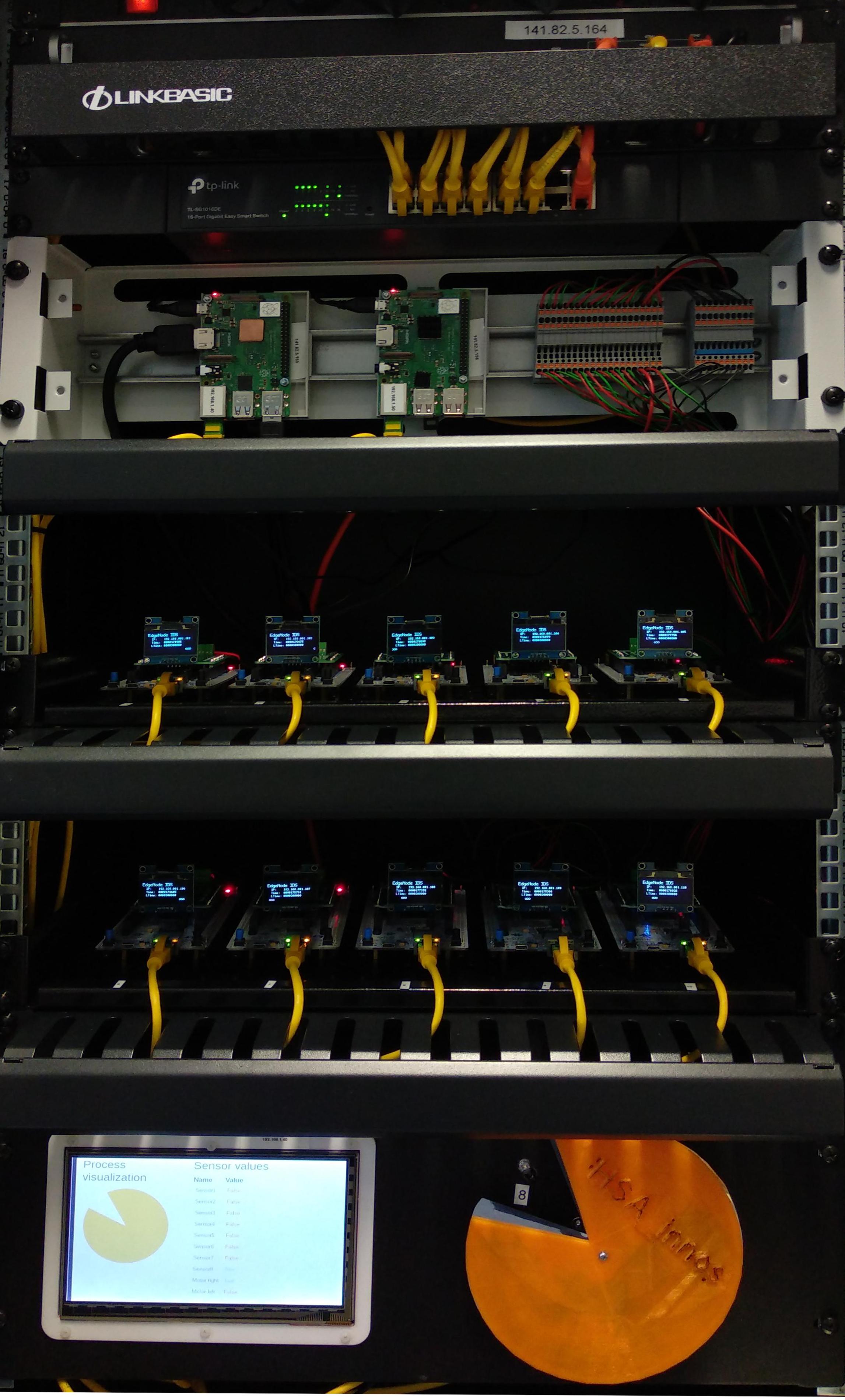}};  
    
\draw[thick,black,decorate,decoration={brace,amplitude=12pt,mirror}] (7.5,3) -- (7.5,8) node[midway, right,xshift=12pt,text width=1cm]{Edge \\ Nodes};   

\draw[thick,black,decorate,decoration={brace,amplitude=12pt,mirror}] (7.5,8.5) -- (7.5,10.0) node[midway, right,xshift=12pt,text width=1cm]{Raspberry Pis};    

\node[align=right,anchor=west] at (7.5,2.5) {HMI};
\draw[->,color=red!50, line width=0.5mm] (7.5,2.5) -- (2.0,1.5);

\node[align=right,anchor=west, text width=1cm] at (7.5,0.5) {Physical \\ process};
\draw[->,color=red!50, line width=0.5mm] (7.5,0.5) -- (5.0,1.2);

\node[align=right,anchor=west] at (7.5,10.5) {NW Switch};
\draw[->,color=red!50, line width=0.5mm] (7.5,10.5) -- (5.0,10.5);

\node[align=right,anchor=west] at (7.5,12.0) {Cloud};
\draw[->,color=red!50, line width=0.5mm] (7.5,12.0) -- (5.0,12.0);
\end{tikzpicture}
\caption{Pictures of the open source \ac{ICS} test-bed, which is controlling a physical process.}
\label{fig:racksetup}
\end{figure}

Figure \ref{fig:racksetup} shows the open-source test-bed for the measurement,
which is set up in a 19"~inch rack.
At the bottom, there is the physical process with a motor controlling a disc.
This disc is sensed by 8 sensors, which are each connected to one edge node.
Additionally, the \ac{HMI} is placed next to the physical process.
In the middle, the edge nodes are placed, each with its own display.
Next to this, the Raspberry Pis are mounted, where one is controlling the \ac{HMI} and the other runs the \ac{PLC} software.
On the top of the rack, the network switch and a server depicting the cloud are placed.

In total, the testbed consists out of 12 Modbus/TCP devices (8 Sensors, 1 Motor, 1 PLC, 1 HMI, 1 SCADA), which are controlling a physical process.

\subsubsection{OpenPLC}
As the central control unit, a Raspberry\footnote{\url{https://www.raspberrypi.org}} Pi with OpenPLC\footnote{\url{https://www.openplcproject.com/}} \cite{alves2014openplc} is used.
The project provides a free open source solution for \acp{PLC}. 
The \ac{PLC} is configured to poll all sensors and actuators every 100\,$ms$. 
Using current sensor states as an input, the program, which is running on the \ac{PLC} is executed and the new output values are calculated.
New values are then written back by the next poll cycle.

\subsubsection{\ac{HMI}}
\acp{HMI} enable technicians to view the current status of a plant and the associated processes. 
Based on this, the technicians can make decisions and interact with the control process.
The \ac{HMI} is a custom implementation based on 
Flask\footnote{\url{http://flask.pocoo.org/}}
with
Pymodbus\footnote{\url{http://riptideio.github.io/pymodbus/}}.
With Pymodbus, the input and output registers are polled from the OpenPLC every 100\,$ms$.

\subsubsection{ScadaBR}
ScadaBR\footnote{\url{http://www.scadabr.com.br/}} is an open source \ac{SCADA} system, which accesses the OpenPLC to gather data.
This is used as a historian and is running on a dedicated server (APU2C4).
The \ac{SCADA} system polls the OpenPLC every 100\,$ms$ via Modbus/TCP.
This makes it possible, on the one hand, to record the data and make it usable for later usage, 
but also to set alerts when values are exceeded or unwanted conditions occur.

\subsubsection{IDS Webserver on the Edge Nodes}
\begin{figure}[htb]
	\centering
	\includegraphics[width=0.55\columnwidth]{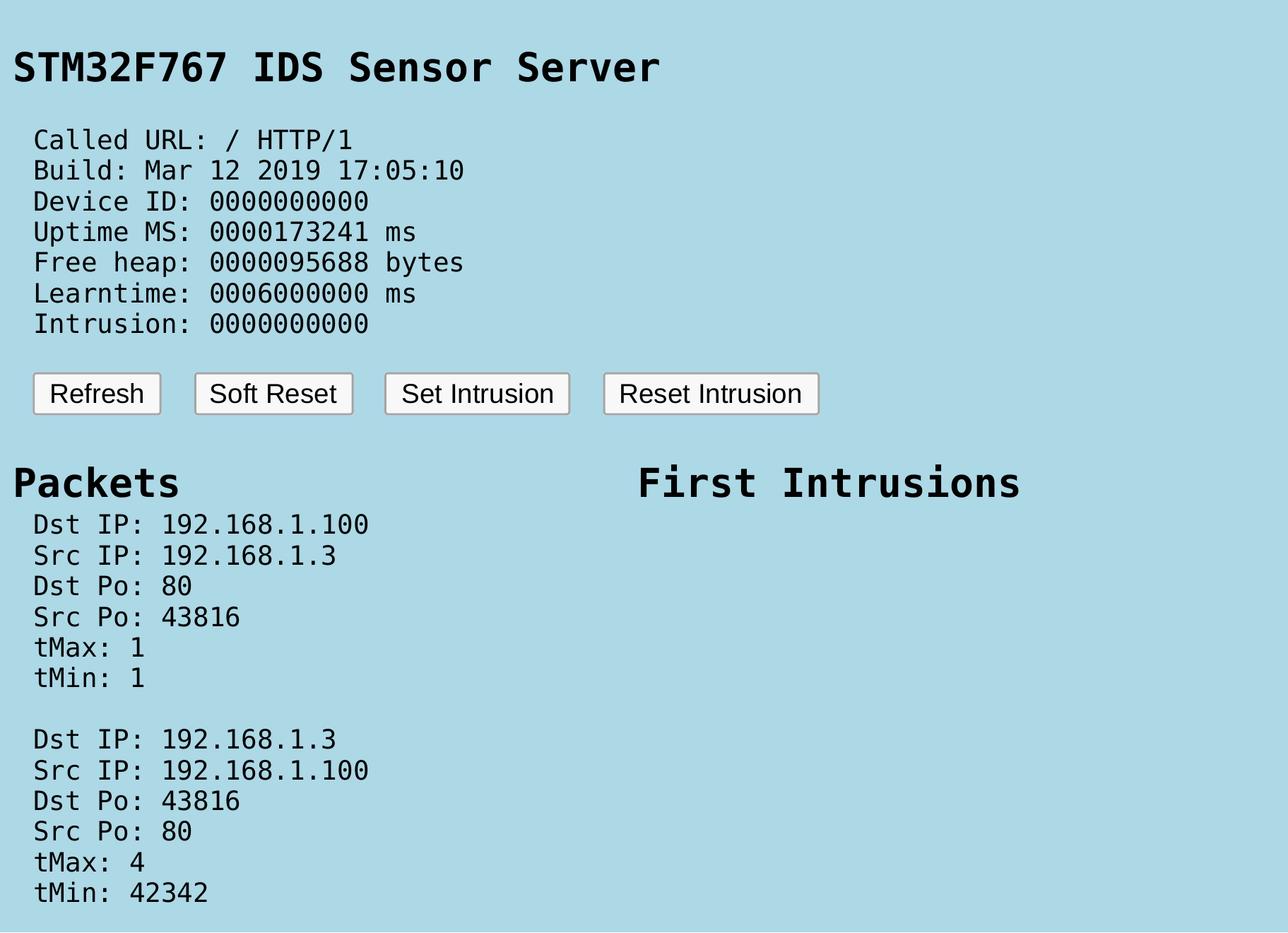}
	\caption{Webpage running on each edge node, displaying the current \ac{IDS} status and debug output.}
	\label{fig:ids_output}
\end{figure}

The webserver enables easy remote maintenance of the edge devices, e.g. from the control room.
\Cref{fig:ids_output} is showing the web page running on each edge node.

At the top the debug information is shown.
To the left, the learned connections are listed.
The background color of the web page displays the current status of the \ac{IDS}: Learning (blue), active without intrusion (green) and intrusion detected (red).
This simple yet effective visualization permits the operator to grasp the network status at a single glance.

\subsubsection{Centralized Logger}
The central logger first checks the \ac{HMAC} signature and whether the message time, which is used as replay attack protection, is consecutive.
\Cref{lst:idslogger} shows the status outputs of the central logger.
In line 1, the edge node device is down, which could be caused e.g.~by a failure or a \ac{DoS} attack.
Line 2 shows the desired status, where the device is up and no intrusion is detected.
An intrusion on edge node 3 is illustrated in line 3.

\begin{lstlisting}[language=bash,basicstyle=\footnotesize, caption=Output of the central logger, label=lst:idslogger]
ID: 1 is <@\textcolor{red}{down}@> Intrusion: <@\textcolor{red}{???}@>
ID: 2 is <@\textcolor{green}{up}@>   Intrusion: <@\textcolor{green}{no}@>
ID: 3 is <@\textcolor{green}{up}@>   Intrusion: <@\textcolor{red}{yes}@>
...
\end{lstlisting}

As explained in \Cref{sec:intrusionannouncement}, the \ac{UDP} broadcast can be received from everywhere in the broadcast domain.
Furthermore, the message outputs can be easily integrated in different logging mechanisms and tools, offering great flexibility to the user.

\subsection{Interarrival Time in the \ac{ICS} Testbed}
In this section we describe, how to statistically assess the deterministic timings in the testbed, which are then utilized by the \ac{IDS} to detect intrusions.

\begin{figure}[htb]
	\centering
	\includegraphics[width=0.85\columnwidth]{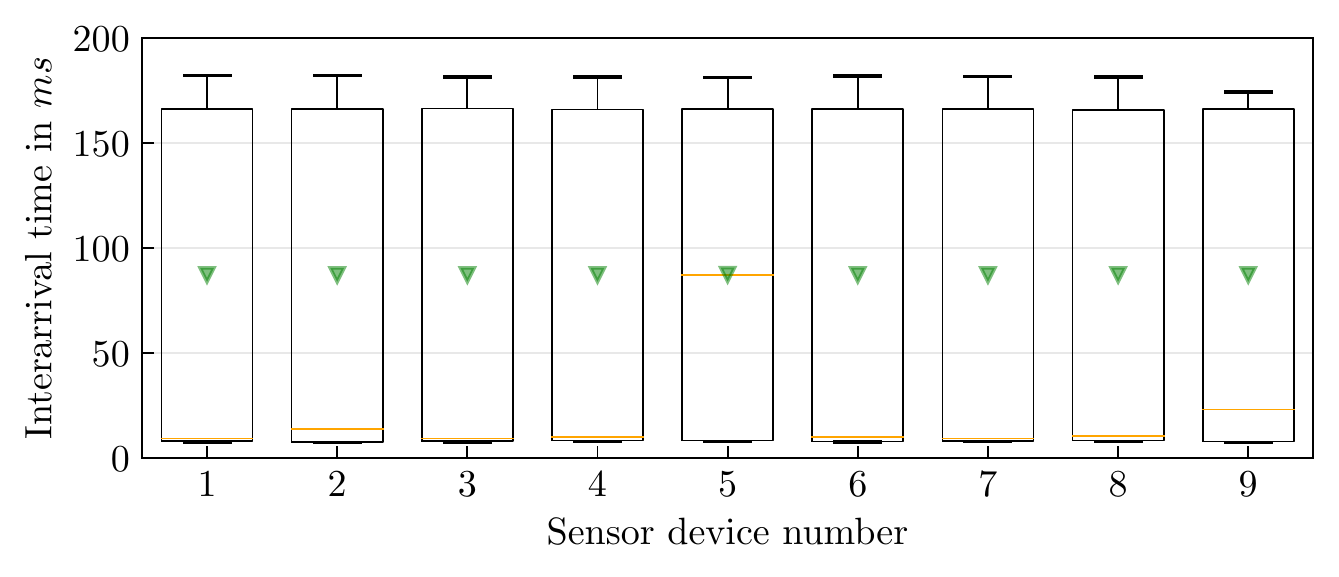}
	\caption{Interarrival time of modbus packets.}
	\label{fig:interarrival_modbus}
\end{figure}

\paragraph{Assessing the packet interarrival time:}
The interarrival time is the time between packets of one connection.
Figure \ref{fig:interarrival_modbus} shows the interarrival time of Modbus/TCP packets to the sensor system in the testbed.
In the box plot, the green arrows are the mean and the orange lines the median.
The plot only contains communication on port 502, respectively Modbus/TCP.
The arithmetic average is about 100\,$ms$ where the Modbus/TCP ``read discrete input'' is executed.
This reflects a normal polling behavior of an industrial \ac{PLC}.
The OpenPLC implementation polls each node.
The timeout is set to 1000\,$ms$ by default.
Most of the controllers have implemented such a timeout which, if exceeded, indicates problems in the network communication.
 
\begin{figure}[htb]
	\centering
	\includegraphics[width=0.85\columnwidth]{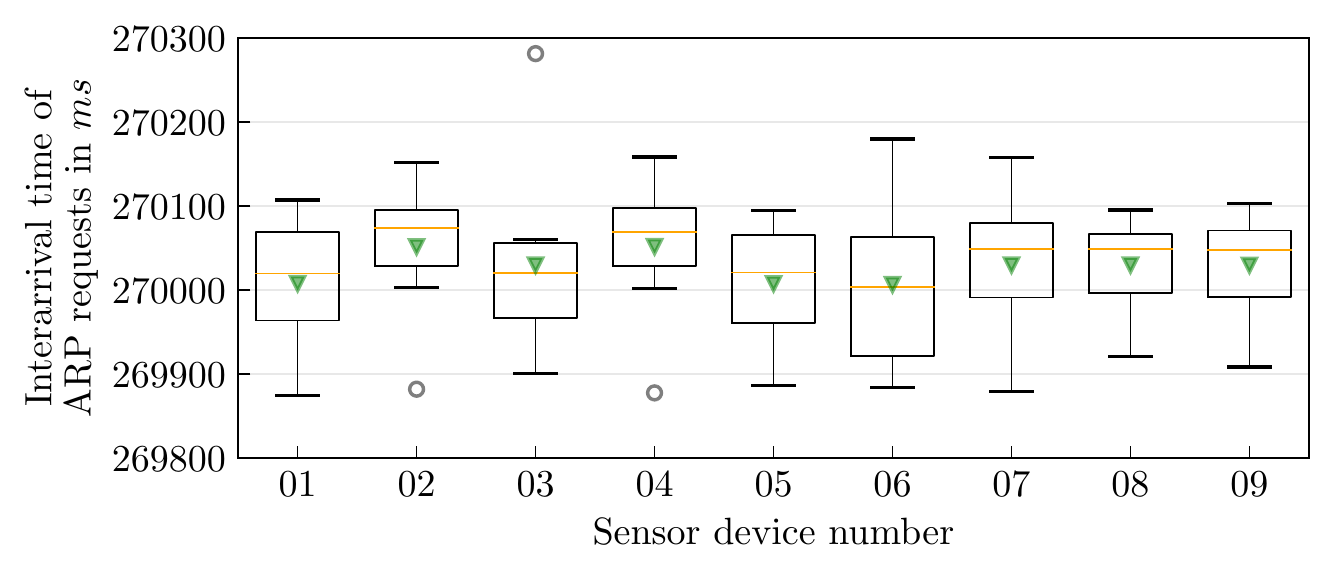}
	\caption{Interarrival time of ARP request packets.}
	\label{fig:interarrival_arp}
\end{figure}

\paragraph{Assessing ARP request interarrival time:}
\Cref{fig:interarrival_arp} illustrates the interarrival time of \ac{ARP} packets in the testbed.
In this figure, all \ac{ARP} requests and responses to the specific sensor are plotted.
This results in a mean of 270 seconds respectively every 4.5 minutes,
when the \ac{ARP} cache is cleared.
This makes it possible to perform a host up detection of devices in the network. 
For example, if there is no \ac{ARP} request for a long time, the device is probably offline.

\begin{figure}[htb]
	\centering
	\includegraphics[width=0.85\textwidth]{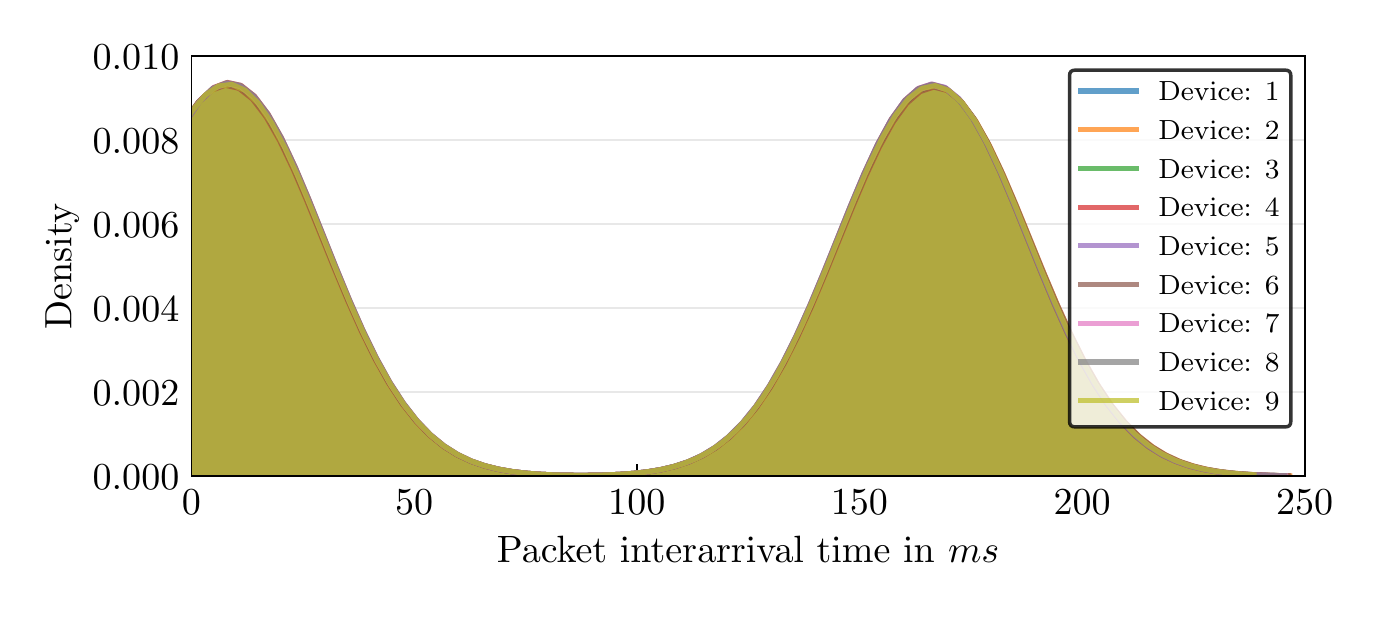}
	\caption{KDE of the interarrival time of Modbus/TCP packets.}
	\label{fig:interarrival_kde_modbus}
\end{figure}

\Cref{fig:interarrival_kde_modbus} shows the \ac{KDE} plot with two high density areas.
In this case, this type of plot represents the frequency of connections over time.
The OpenPLC v3 implementation uses a default polling sleep time of 100\,$ms$,
which is added on top of the interaction of each node.
The first peak is generated, because of the default response time of a query.
This first peak represents the request from the \ac{PLC} and the fastest possible answer by the sensor (request $\rightarrow$ response).
The second peak is the delay to not flood the edge nodes, with the default of 100\,$ms$.
This is, in other words, the refresh rate of the \ac{PLC} values.

All Modbus/TCP interarrival times of the nine edge nodes are actually shown in the plot, 
but they overlap so much, that almost no difference is visible. 
This could be different, e.g.~if two \acp{PLC} poll the sensors.
However, this does not make any difference in the classification per connection and does not change the core statement of this plot.

\subsection{Intrusion Detection Benchmark}
The parameters used to detect attacks essentially depend on what has been learned 
and how the attack is executed.

\subsubsection{Attack Detection}
The introduced attacker models in \Cref{subsec:attackermodel} are used to measure the attack detection of the \ac{IDS} on the embedded edge nodes.
The following scenarios are covered by these two attacker models,
which are summarized in \Cref{tab:attacks_sumamry_detection}:

The local attacker \textbf{removes} \circled{\color{white}1} an edge node from the network.
The centralized logger will detect this incident after some seconds, because the keep alive message of the edge node is missing. 

The network is \textbf{actively sniffed} \circled{\color{white}2} by the local attacker with e.g. \ac{ARP} poisoning.
Caused by the \ac{ARP} poisoning, \ac{ARP} requests are sent out. 
Those packets are new to the \ac{IDS}, which results in an incident report.
Since this attack already requires some knowledge, the attacker knowledge is categorized as medium.

Input, output or keep alive commands are \textbf{spoofed} \circled{\color{white}3} by the local attacker.
If an attacker connects his own device to the network and sends out \ac{ARP} requests,
the edge node will receive those requests from an unknown source and will report this.

The local or remote attacker \textbf{inject} \circled{\color{white}4} packets into the network to e.g. to send commands to the edge nodes.
New connections will be detected by the \ac{IDS}.

An attacker is flooding an edge node to perform a \textbf{\ac{DoS} attack} \circled{\color{white}5}.
The edge node \ac{IDS} detects, that packets are either new or occur too often.
Additionally, this will be detected by the logging server, because no alive message are received anymore.

The attacker is \textbf{passively sniffing} \circled{\color{white}6} the network with a unidirectional network diode.
If this is done completely passive, the \ac{IDS} can not recognize this.
For this kind of attack, the attacker knowledge is low.

An attack is already executed during the \textbf{learning} \circled{\color{white}7} phase of the \ac{IDS}.
Of course, if an attacker manages to get involved during the learning phase, the IDS naturally also learns it and regards this as regular.
However, the attacker must continue this traffic after the learning phase, otherwise the 
\ac{IDS} will detect the attack.

The attacker \textbf{captures} \circled{\color{white}8} an edge node.
The \ac{IDS} is only capable to detect the attack during the attack phase.
After an successful attack and if the edge node is captured, the attacker can manipulate the data on the captured edge node.
If the attacker opens new connections, the other still trusted edge nodes will detect this.

\begin{table}[htb]
    \centering
    \caption{Summary of the evaluated attack scenarios and detection capabilities.}
    \label{tab:attacks_sumamry_detection}
        \begin{tabular}{l l l l}
            \hline
            \textbf{Model} & \textbf{Short description} & \textbf{Attacker} & \textbf{Detection} \\
            \hline
            \circled{\color{white}1}   & Node removed      & weak           & \cmark \\
            \circled{\color{white}2}   & Active sniffing   & medium         & \cmark \\
            \circled{\color{white}3}   & Spoofing attack   & medium         & \cmark \\
            \circled{\color{white}4}   & Injection attack  & weak           & \cmark \\
            \circled{\color{white}5}   & \ac{DoS} attack   & weak           & \cmark \\
            \circled{\color{white}6}   & Passive sniffing  & weak           & \xmark \\
            \circled{\color{white}7}   & Learning attack   & strong         & \omark \\
            \circled{\color{white}8}   & Capture edge node & strong         & \omark \\
            \hline
            \multicolumn{4}{r}{\cmark detected \omark dependent \xmark not detected}
        \end{tabular}
\end{table}

\subsubsection{Time for Learning Regular Behavior}
The time needed by the \ac{IDS} to learn regular connections depends first of all on the use case.
In the case of our open source testbed, 
there is normally the distinction between direct connections, 
such as the read and write of the register through the \ac{PLC} 
and packets like \ac{ARP} requests that the edge node receives,
because it is in the broadcast domain.
In our testbed, the time for learning is approximately two times the time between the longest interarrival time of an \ac{ARP} request, which is approximately 10 minutes.
This is the absolute minimum and must always be chosen to collect rarer events. 
In practice, it will make no difference whether the initial learning time is a few minutes, hours or even one day, as long as the devices are not battery operated.

\subsection{\ac{MCU} Performance Data}

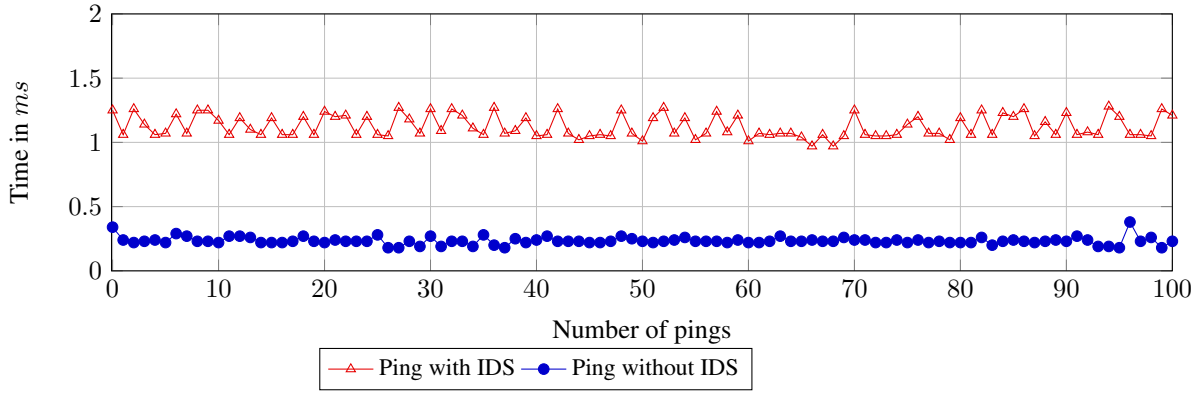
\begin{figure}[htb]
	\centering
	\begin{tikzpicture}
	\begin{axis}[
	width=0.95\columnwidth,
	height=5cm,
	xlabel={Number of pings},
	ylabel={Time in $ms$},
	ymin = -0.0,
	xmin = -0.1,
 	ymax = 2,
	xmax = 100,
	grid=both,
	grid style={line width=.1pt, draw=gray!10},
	major grid style={line width=.2pt,draw=gray!50},
	legend style={at={(0.4,-0.3)},anchor=north},
	legend columns=4,
	]
	\addplot+[red!90!black,mark=triangle,mark options={fill=red!90!black}] table [x=time,y=ids,col sep=comma] {ping.csv};
	\addlegendentry{\small Ping with IDS};
	\addplot+[blue!80!black,mark=*,mark options={fill=blue!80!black}] table [x=time,y=without,col sep=comma] {ping.csv};
	\addlegendentry{\small Ping without IDS};
	\end{axis}
	\end{tikzpicture}
	\caption{Measurement of the ping behavior with and without the IDS}
	\label{fig_ping}
\end{figure}

\Cref{fig_ping} shows the ping of the edge node device with enabled and disabled \ac{IDS}.
This is measured with fping\footnote{\url{https://fping.org/}} every 100\,$ms$ over 100 samples.
The average ping without the \ac{IDS} is about 0.31\,$ms$ and 1.13\,$ms$ with the enabled \ac{IDS}.

The data throughput measured with iperf drops from 28.2\,$Mbits/sec$ without the \ac{IDS} to 4.23\,$Mbits/sec$ with the \ac{IDS}.
However, in our testbed, this performance reduction does not have any influence on the controlled process, because there are only about 100\,$Kbits/sec$ with a \ac{PLC} cyclic refresh time of 50\,$ms$.
Such low traffic rates are common in industrial plants, but also with higher rates the \ac{IDS} is capable to handle the traffic.
 
\Cref{tab:build} shows the differences in the software build with \ac{IDS} and whithout \ac{IDS}.
It is only the function analyzing the received and transmitted packets, because the other functions depend on the configuration.
This means, for example, the webserver or the logging function can consume more or less RAM and ROM depending on the configuration.
Due to this reason, the comparison is limited to the analyze function of the \ac{IDS}.
        
\begin{table}[htb]
    \centering
    \caption{Binary comparison of example application with and without the \ac{IDS} building block.}
    \label{tab:build}
        \begin{tabular}{l l l l l l}
            \hline
            \textbf{Information} & \textbf{text} & \textbf{data} & \textbf{bss} & \textbf{dec} & \textbf{hex} \\
            \hline
            With \ac{IDS}    & 145176 & 12592 & 285848 & 443616 & 6c4e0 \\
            Without \ac{IDS} & 141872 & 12592 & 285832 & 440296 & 6b7e8 \\
            \hline
            Difference      & 3304  & 0     & 16    & 3320  & cf8  \\
        \end{tabular}
\end{table}

This analysis shows, that the actual \ac{IDS} functionality requires quite few resources and thus does not waste scarce flash memory.
          
\section{Conclusion \label{sec:conclusion}}
Our work shows, that network based \acp{IDS} on low"~performance \acp{MCU} are a feasible and potent way of detecting intrusions in industrial networks.
Especially the periodical nature of polled communication between \acp{PLC} and remote sensors and actors allow effective detection mechanisms.
The measurements conducted, that numerous network"~based cyber attacks are detected reliably, while the detection itself affects the network traffic only slightly.
Our \ac{IDS} approach can be used either as a single point of defense or can be easily combined with e.g.~firewalls.
Combined with the protocol neutrality of the metadata approach proposed in this paper,
the distributed \ac{IDS} can be effortlessly integrated within already existing projects as an autonomous level of security.
To base the implementation of the \ac{IDS} as a function on the widely used \ac{LwIP} stack 
provides a decent foundation to upgrade the system to an \ac{IPS}.
This is possible, because as an intrusion detected packet can be dropped before it is being processed
by the regular application running on the \ac{MCU}.
The good detection results paired with the modularity and easy"~to"~integrate nature of the proposed approach make it
a reasonable fitting choice to tackle the upcoming security problems of the Industry 4.0.

\bibliographystyle{plain}
\bibliography{\jobname}

\begin{acronym}
  \acro{A}{Availability}
  \acro{ACK}{acknowledgment}
  \acro{API}{Application Programming Interface}
  \acro{ARP}{Address Resolution Protocol}
  \acro{BMBF}{Federal Ministry of Education and Research}
  \acro{C}{Confidentiality}
  \acro{CIA}{Confidentiality, Integrity and Availability}
  \acro{CID}{Company IDentifier}
  \acro{CPE}{Common Platform Enumeration}
  \acro{CPS}{Cyber Physical System}
  \acro{CPU}{Central Processing Unit}
  \acro{CRT}{Communication Robustness Test}
  \acro{CSV}{Comma-Separated Values}
  \acro{CVE}{Common Vulnerabilities and Exposures}
  \acro{DCS}{Distributed Control System}
  \acrodefplural{DCS}[DCSs]{Distributed Control Systems}
  \acro{DHCP}{Dynamic Host Configuration Protocol}
  \acro{DoS}{Denial of Service}
  \acro{DuT}{Device under Test}
  \acrodefplural{DuT}[DuTs]{Devices under Test}
  \acro{ERP}{Enterprise Resource Planning}
  \acro{FTP}{File Transfer Protocol}
  \acro{HAL}{Hardware Abstraction Layer}
  \acro{HMAC}{Hash Message Authentication Code}
  \acro{HTTP}{Hypertext Transfer Protocol}
  \acro{HTTPS}{Hypertext Transfer Protocol Secure}
  \acro{HMI}{Human Machine Interface}
  \acrodefplural{HMI}[HMIs]{Human Machine Interfaces}
  \acro{I}{Integrity}
  \acro{IEEE}{Institute of Electrical and Electronics Engineers}
  \acro{ICS}{Industrial Control System}
  \acrodefplural{ICS}[ICSs]{Industrial Control Systems}
  \acro{IDE}{Integrated Development Environment}
  \acro{IPS}{Intrusion Prevention System}
  \acrodefplural{IPS}[IPSs]{Intrusion Prevention Systems}
  \acro{IDS}{Intrusion Detection System}
  \acrodefplural{IDS}[IDSs]{Intrusion Detection Systems}
  \acro{IoT}{Internet of Things}
  \acro{IO}{Input/Output}
  \acro{IIoT}{Industrial Internet of Things}
  \acro{IP}{Internet Protocol}
  \acro{I2C}{Inter-Integrated Circuit}
  \acro{KDE}{Kernel Density Estimation}
  \acro{LED}{Light-emitting Diode}
  \acro{LwIP}{Lightweight IP}
  \acro{M2M}{Machine to Machine}
  \acro{MAC}{Media Access Control}
  \acro{MCU}{Micro Controller Unit}
  \acrodefplural{MCU}[MCUs]{Micro Controller Units}
  \acro{MES}{Manufacturing Execution System}
  \acro{MitM}{Man-in-the-Middle}
  \acro{NSE}{Nmap Scripting Engine}
  \acro{OS}{Operating System}
  \acro{OSI}{Open Systems Interconnection}
  \acro{OT}{Operational Technology}
  \acro{OUI}{Organizationally Unique Identifier}
  \acro{pcap}{packet capture}
  \acrodefplural{pcap}[pcaps]{packet captures}
  \acro{PCB}{Printed Circuit Board}
  \acro{PLC}{Programmable Logic Controller}
  \acrodefplural{PLCs}{Programmable Logic Controllers}
  \acro{PoC}{Proof of Concept}
  \acro{PSK}{Pre-Shared Key}
  \acro{RA}{Registration Authority}
  \acro{SCADA}{Supervisory Control and Data Acquisition}
  \acro{SSH}{Secure Shell}
  \acro{SYN}{synchronize}
  \acro{TAP}{Terminal Access Point}
  \acro{TCP}{Transmission Control Protocol}
  \acro{ToS}{Type of Service}
  \acro{TTL}{Time to Live}
  \acro{UDP}{User Datagram Protocol}
  \acro{USB}{Universal Serial Bus}
  \acro{VLAN}{Virtual Local Area Network}
  \acrodefplural{VLAN}[VLANs]{Virtual Local Area Networks}
  \acro{VM}{Virtual Machine}
  \acro{VPN}{Virtual Private Network}
\end{acronym}

\end{document}